\documentclass{appolb}
\usepackage{graphicx}
\usepackage{color}
\usepackage{amsmath} 
\usepackage{slashed}
\usepackage{cite}
\newcommand{\MSb}{$\overline{\mbox{MS}}$}
\begin{document}

\noindent
{\small DESY 21-209 \hfill November 2021} \\
\title{Renormalization of non-singlet quark operator matrix elements for deep-inelastic scattering%
\thanks{Presented at Matter To The Deepest 2021.}%
}
\author{Sam Van Thurenhout, Sven-Olaf Moch
\address{II.~Institute for Theoretical Physics, Hamburg University\\
\vspace{0.1cm}
D-22761 Hamburg, Germany}
}
%\maketitle
{\let\newpage\relax\maketitle}
\begin{abstract}
We introduce a new method for calculating the mixing matrix for non-singlet quark operators including total derivatives, based solely on their renormalization structure in the chiral limit. As input, the method requires the well-known forward anomalous dimensions, which determine the evolution of parton distribution functions, and a calculation of the matrix elements of operators without total derivatives. Assuming a large number of quark flavors $n_f$, we are able to calculate the mixing matrix to fifth order in the strong coupling $\alpha_s$ in the \MSb-scheme.
\end{abstract}
  
\section{Introduction}
The physics of the strong interaction is accurately described by a gauge theory based on $SU(3)$, namely quantum chromodynamics (QCD). While QCD has been very successful, some open questions remain. An important example of such an as-of-yet unanswered question is the following: \textit{How do hadronic properties emerge from the properties of the constituent partons?} For example, it is still not completely understood how the spin of the proton arises from the angular momenta of the quarks and gluons inside the proton. Hence we need to look inside the proton, which of course we can do using scattering experiments. The description of such experiments in QCD is somewhat simplified because of factorization, which tells us that physical cross sections can be written as a convolution of a hard-scale function and a soft-scale one. The hard-scale function represents the short-distance part of the process, which involves the partonic degrees of freedom inside the hadron. Typically it can be written as some partonic cross section, calculable perturbatively in the strong coupling $\alpha_s$. The soft-scale function represents the long-distance part of the process and is non-perturbative. This means that it has to be fitted directly from experiment or calculated using non-perturbative techniques, e.g. using lattice QCD. Important examples of such non-perturbative functions are the parton distribution functions (PDFs) and the generalized parton distributions (GPDs).

PDFs are accessible in forward-kinematic processes like inclusive deep-inelastic scattering (DIS), $e p \rightarrow e X$, and describe the distribution of the longitudinal momentum and polarization of partons inside hadrons. From the experimental side, they can be accessed by analyzing the data from the HERA collider~\cite{Abramowicz:2015mha,Accardi:2016ndt} and the planned Electron Ion Collider (EIC)~\cite{Boer:2011fh,AbdulKhalek:2021gbh} in the future. GPDs can be considered to be the counterparts of the standard PDFs in processes with off-forward kinematics, like e.g. exclusive deeply-virtual Compton scattering (DVCS, $e p \rightarrow e p \gamma$). They describe the transverse distributions of the partons in the hadronic target. Combining this with the longitudinal information then gives rise to a full three-dimensional description of hadronic structure. Furthermore, GPDs also allow for the determination of the partonic angular momentum contributions to the total hadronic spin~\cite{Diehl:2003ny}. The study of GPDs is one of the main goals of the future EIC~\cite{Boer:2011fh,AbdulKhalek:2021gbh}. The properties of PDFs and GPDs can also be studied using lattice QCD, see e.g.~\cite{Gockeler:2004wp, Gockeler:2010yr} 
and for recent progress~\cite{Braun:2015axa,Braun:2016wnx,Bali:2018zgl,Bali:2019dqc,Harris:2019bih,Alexandrou:2020sml}.

While the distribution functions themselves are non-perturbative, their scale-dependence can be calculated perturbatively in the strong coupling $\alpha_s$. The origin of this lies in the analysis of DIS and DVCS using the Wilsonian operator product expansion (OPE)~\cite{Wilson:1969zs, Zimmermann:1972tv}, which gives a direct relation between PDFs and GPDs and the matrix elements of composite local gauge invariant operators. The consequence of this is that the scale-dependence of the parton distributions is directly related to the scale-dependence of the operators, determined by their anomalous dimensions. Hence it is important to understand the renormalization properties of the relevant operators, both in forward and off-forward kinematics. In this work we focus on the leading-twist flavor-non-singlet quark operators.

In forward kinematics, the anomalous dimensions of the twist-2 non-singlet quark operators, written as functions of the Mellin moment $N$ representing their Lorentz spin, are known completely up to the three-loop level~\cite{Gross:1973ju, Floratos:1977au, Moch:2004pa,Blumlein:2021enk}. In certain limits, partial information is also available at four and five loops~\cite{Velizhanin:2011es,Velizhanin:2014fua,Ruijl:2016pkm,Moch:2017uml,Herzog:2018kwj}. In off-forward kinematics, the evolution kernel for the non-singlet operators is known completely up to three loops~\cite{Braun:2017cih}. The calculation exploited conformal symmetry~\cite{Braun:2003rp} of the QCD Lagrangian near the Wilson-Fisher fixed point. This technique was already introduced in the nineties by M\"uller and Belitsky~\cite{Mueller:1993hg,Belitsky:1998gc} for the calculation of two-loop radiative corrections. In addition to the usual variable $N$, the moment space anomalous dimensions in off-forward kinematics also depend on the number $k$ of total derivatives acting on the operator. Besides the computations using the conformal approach, there are also some fixed-moment calculations of the operator matrix elements (OMEs) up to the three-loop level. These calculations were done in the modified minimal subtraction (\MSb) scheme as well as in alternative ones, like the regularization invariant (RI) scheme~\cite{Gracey:2009da,Kniehl:2020nhw}. An advantage of such schemes is that they are suitable for direct application to available lattice QCD results. These fixed-moment calculations use a different basis for total derivative operators from the one in the conformal approach, making a direct comparison between the fixed-moment results of~\cite{Gracey:2009da,Kniehl:2020nhw} and~\cite{Braun:2017cih,Mueller:1993hg,Belitsky:1998gc} impossible. Instead, additional computational steps are required.

In the present article, we will review the renormalization of flavor-non-singlet quark operators including total derivatives, paying particular attention to possible choices for their bases. This way we can connect different results which, as of yet, appeared unrelated in the literature. Assuming a large number of quark flavors $n_f$, we calculate the relevant OMEs up to four-loop order for a non-zero momentum transfer through the operator vertex. Furthermore, we derive consistency relations for the corresponding operator anomalous dimensions, which allow us to check and extend previous calculations for the leading-$n_f$ terms of the off-forward anomalous dimensions up to five loops.

These proceedings are organized as follows. Next, we define the operators and their matrix elements and study their renormalization. In Section \ref{sec:bases}, we then discuss two bases for total derivative operators used in the literature, and summarize the knowledge of the mixing matrices in these bases. The next section introduces a consistency relation which the anomalous dimensions have to obey, leading to a novel algorithm for deriving the full mixing matrix. Results of the application of this algorithm are discussed in Section \ref{sec:results}, and we finish with some concluding remarks in Section \ref{sec:conclusion}.
\label{sec:renorm}
\section{Operator renormalization}
\subsection{Operators and their matrix elements}
\begin{figure}[htb]
\centerline{%
\includegraphics[width=0.5\textwidth]{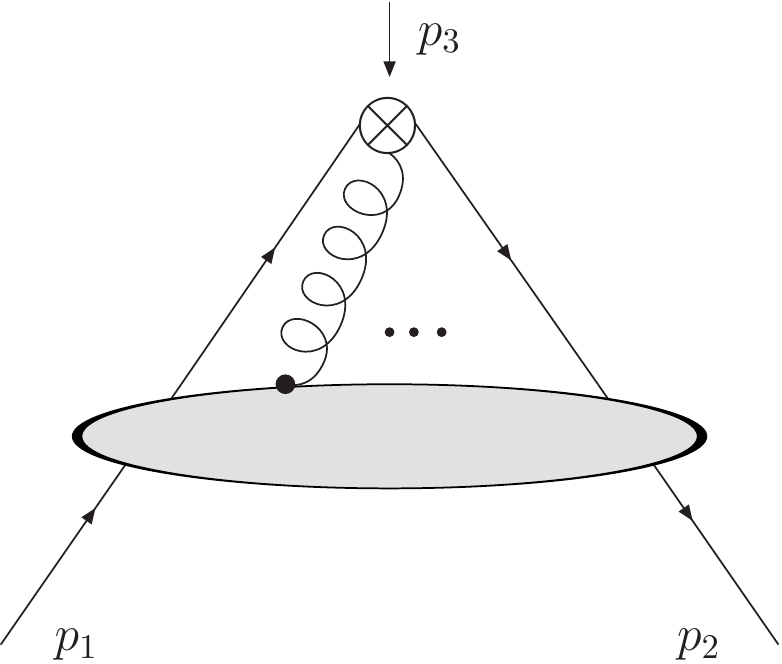}}
\caption{The Green's function 
$\langle \psi(p_1) | O_{\mu_1\dots \mu_N}^{NS}(p_3) | \overline{\psi}(p_2)\rangle$ 
with momentum $p_3=-p_1-p_2$ flowing through the operator vertex. 
For simplicity, we set $p_2 \equiv 0$.}
\label{figGreenFun}
\end{figure}

The operators appearing in the OPE analysis of DIS and DVCS are the spin-$N$ local non-singlet quark operators
\begin{equation}
\label{eq:OpDef}
    \mathcal{O}^{NS}_{\mu_1 \dots \mu_{N}} \,=\, 
    \mathcal{S}\, \overline{\psi}\lambda^{\alpha}\gamma_{\mu_1} D_{\mu_2} \dots D_{\mu_{N}}\psi\, ,
\end{equation}
with $\psi$ the quark field, $D_{\mu} = \partial_{\mu} - i g_s A_{\mu}$
the standard QCD covariant derivative, and $\lambda^{\alpha}$ the generators
of the flavor group $SU(n_f)$. 
As we are interested in the leading-twist contributions, 
we symmetrize the Lorentz indices and take the traceless part, indicated by ${\mathcal S}$. 
{This projects the twist-two contribution, see e.g.~\cite{Blumlein:1999sc}.} 
The scale-dependence of these operators is determined by their anomalous dimensions, which can be calculated perturbatively in the strong coupling. Schematically
\begin{equation}
    \frac{d \mathcal{O}}{d \ln \mu^2} = -\gamma \mathcal{O}, \: \: \gamma \equiv a_s \gamma^{(0)} +a_s^2 \gamma^{(1)} + ...
\end{equation}
with $\mu$ the renormalization scale and $a_s = \alpha_s/(4\pi)$. We gain access to the anomalous dimensions by considering spin-averaged matrix elements of the operators, 
\begin{equation}
\label{eq:generalOME}
    \langle \psi(p_1) | \mathcal{O}_{\mu_1 \dots \mu_N}^{NS}(p_3) | \overline{\psi}(p_2)\rangle 
    \,,
\end{equation}
with quarks and anti-quarks of momenta $p_1$ and $p_2$ as external fields, see Fig.\ref{figGreenFun}. We assume all momenta to be incoming, $\sum_{i=1}^3 p_i = 0$.
The tracelessness and symmetry of the Lorentz indices is most easily achieved by contracting the OMEs with a tensor of light-like $\Delta$,
\begin{equation}
    \label{eq:theDeltas}
    \Delta^{\mu_1}\dots \Delta^{\mu_N}
    \, ,
\end{equation}
and $\Delta^2=0$. As we are interested in the renormalization of the non-singlet operators including total derivatives, we have to choose $p_3\neq0$. However, for simplicity but without loss of generality, one can nullify one of the external momenta, i.e. we can set $p_2=0$. The calculated OMEs are then of the form
\begin{equation}
\label{eq:OMEs}
    \Delta^{\mu_1}\dots \Delta^{\mu_N}\, \langle \psi(p_1) | \mathcal{O}_{\mu_1\dots \mu_N}^{NS}(-p_1) | \overline{\psi}(0) \rangle \equiv \langle \psi(p_1) | \mathcal{O}_{N}(-p_1) | \overline{\psi}(0) \rangle
    \, .
\end{equation}
Note that this reduces the initial three-point function to a two-point one. The computation of these OMEs is done entirely automatically using the {\sc Form}~\cite{Vermaseren:2000nd,Kuipers:2012rf} program {\sc Forcer}~\cite{Ruijl:2017cxj}, resulting in fixed moments of the OMEs in Eq.~(\ref{eq:OMEs}).

\subsection{Operator renormalization and anomalous dimensions}

The actual renormalization will be done using the \MSb-scheme \cite{tHooft:1973mfk, Bardeen:1978yd}, in which the evolution of the strong coupling is governed by
\begin{equation}
    \frac{da_s}{d\ln \mu^2} = \beta(a_s) =-a_s(\epsilon+\beta_0 a_s +\beta_1 a_s^2 +\beta_2 a_s^3 + \dots).
\end{equation}
Here $\beta(a_s)$ is the standard QCD beta-function with $\beta_0 = (11/3) C_A - (2/3) n_f$. $C_A$ is the quadratic Casimir of the adjoint representation of the color group $SU(n_c)$, $C_A = n_c$.

In forward kinematics the operators renormalize multiplicatively as
\begin{equation}
\label{eq:forRen}
        \mathcal{O}_{N+1} = Z_{N,N}[\mathcal{O}_{N+1}].
    \end{equation}
The corresponding anomalous dimensions are related to the QCD splitting functions by a Mellin transform
\begin{equation}
    \gamma_{NS}(N) \equiv \gamma_{N-1,N-1} = -\int_{0}^{1}\text{d}x \: x^{N-1}{P_{NS}(x)}
\end{equation}
which determine, through a convolution $\otimes$ defined as
\begin{equation}
    [P_{NS}\otimes f_{NS}](x) = \int_x^1 \frac{dy}{y} P_{NS}(y)f_{NS}\Big(\frac{x}{y}\Big),
\end{equation}
the scale-dependence of the standard PDFs
\begin{equation}
    \frac{\text{d} f_{NS}(x,\mu^2)}{\text{d} \ln{\mu^2}} = [P_{NS} \otimes f_{NS}](x).
\end{equation}
 This is the well-known DGLAP evolution equation \cite{Gribov:1972ri, Altarelli:1977zs, Dokshitzer:1977sg}.
For off-forward kinematics, the operator renormalization becomes more complicated because of mixing with total derivative operators. This means that now the renormalization takes the form of a matrix equation
\begin{equation}
        \begin{pmatrix}
            \mathcal{O}_{N+1} \\ \partial \mathcal{O}_{N}\\ \vdots \\ \partial^N \mathcal{O}_{1}
    \end{pmatrix} 
    \,=\, 
    \begin{pmatrix}
            Z_{N,N} & Z_{N,N-1} & ... & Z_{N,0} \\
            0 & Z_{N-1,N-1} & ... & Z_{N-1,0} \\
            \vdots & \vdots & ... & \vdots  \\
            0 & 0 & ...  &  Z_{0,0}
    \end{pmatrix} 
    \begin{pmatrix}
            [\mathcal{O}_{N+1}] \\ [\partial \mathcal{O}_{N}] \\ \vdots 
             \\ [\partial^N \mathcal{O}_{1}]
    \end{pmatrix}.
    \end{equation}
The off-forward anomalous dimensions, which determine the GPD scale-dependence \cite{Diehl:2003ny}, are related to the $Z$-factors by
\begin{equation}
            \gamma_{N,k}^{\mathcal{D}} \,=\,
-\,\bigg( \,\frac{d }{d\ln\mu^2 }\; Z_{N,j} \bigg)\, Z_{j,k}^{\,-1}
        \end{equation}
and can be expanded in a power series in the strong coupling
\begin{equation}
    \gamma_{N,k}^{\mathcal{D}} \,=\, 
    a_s\gamma_{N,k}^{\mathcal{D},(0)} + a_s^2\gamma_{N,k}^{\mathcal{D},(1)} + a_s^3\gamma_{N,k}^{\mathcal{D},(2)} + a_s^4\gamma_{N,k}^{\mathcal{D},(3)} + a_s^5\gamma_{N,k}^{\mathcal{D},(4)}
    + \dots \, .
\end{equation}

\section{Two possible bases for total derivative operators}
\label{sec:bases}
To study the renormalization of the quark operators in off-forward kinematics, we now have to choose a basis for the total derivative operators. In this section, we discuss two possibilities which have appeared in the literature.
\subsection{The Gegenbauer basis}
One approach is to expand the local operators in terms of Gegenbauer polynomials~\cite{Braun:2017cih}
\begin{equation}
        \mathcal{O}_{N,k}^{\mathcal{G}} = (\Delta \cdot \partial)^k \overline{\psi}(x)  \slashed \Delta C_N^{3/2}\Bigg(\frac{\stackrel{\leftarrow}{D} \cdot \Delta-\Delta \cdot \stackrel{\rightarrow}{D}}{\stackrel{\leftarrow}{\partial} \cdot \Delta+\Delta \cdot \stackrel{\rightarrow}{\partial}}\Bigg)\psi(x)
    \end{equation}
where \cite{olver10}
\begin{equation}
        C_N^{\nu}(z) = \frac{\Gamma(\nu+1/2)}{\Gamma(2\nu)}\, \sum\limits_{l=0}^{N}\,
    (-1)^l\binom{N}{l}\frac{(N+l+2)!}{(l+1)!}\, 
    \Big(\frac{1}{2}-\frac{z}{2}\Big)^l.
    \end{equation}
Here $k \geq N$ is the total number of derivatives and 
we use the superscript $\mathcal{G}$ to denote the operators in the Gegenbauer basis. Using properties of the Gamma function we can rewrite the operators as a particular double sum of left- and right-derivative operators
\begin{align}
        \mathcal{O}_{N,k}^{\mathcal{G}} =& \: \frac{1}{2N!}\sum_{l=0}^{N}(-1)^l\binom{N}{l}\frac{(N+l+2)!}{(l+1)!} \nonumber \\& \times \sum_{j=0}^{k-l}\binom{k-l}{j}\overline{\psi}(x)  \slashed \Delta (\stackrel{\leftarrow}{D} \cdot \Delta)^{k-l-j}(\Delta \cdot \stackrel{\rightarrow}{D})^{l+j}\psi(x).
    \end{align}
The Gegenbauer basis is a natural choice when there is conformal symmetry, e.g. near the Wilson-Fisher critical point of QCD, where $\beta_{\text{QCD}} =~0$~\cite{Efremov:1979qk, Belitsky:1998gc, Braun:2017cih}. The anomalous dimension matrix in this basis is triangular, i.e. its elements $\gamma_{N,j}^{\mathcal{G}} = 0$ if $j>N$, and its diagonal elements correspond to the standard forward anomalous dimensions $\gamma_{N,N}$~\cite{Moch:2017uml}, cf. Eq.~(\ref{eq:forRen}). Note that we can drop the superscript ${\mathcal{G}}$ for $\gamma_{N,N}$ as they do not depend on the basis choice for operators with additional total derivatives. Currently the Gegenbauer mixing matrix is known completely to three loops~\cite{Braun:2017cih}.
\subsection{The total derivative basis}
Another approach is to identify the operators by counting powers of derivatives
\begin{equation}
        \mathcal{O}_{{ p},{ q},{r}}^{\mathcal{D}} = (\Delta \cdot \partial)^{{ p}} \Big\{(\Delta \cdot D)^{{ q}}\overline{\psi}\, \slashed \Delta 
    (\Delta \cdot D)^{{r}}\psi\Big\},
    \end{equation}
see e.g.~\cite{Gracey:2011zn,Gracey:2011zg} 
{and \cite{Geyer:1982fk,Blumlein:1999sc}}. The superscript $\mathcal{D}$ indicates that the operators are written in the 
total derivative basis. If we now impose the chiral limit, i.e. work with massless quarks, the partial derivatives act as
\begin{equation}
\label{eq:partialAct}
    \mathcal{O}_{p,q,r}^{\mathcal{D}} \,=\, \mathcal{O}_{p-1,q+1,r}^{\mathcal{D}} + \mathcal{O}_{p-1,q,r+1}^{\mathcal{D}}
    \, .
\end{equation}
Another consequence of the chiral limit is that left- and right-derivative
operators renormalize with the same renormalization constants
\begin{eqnarray}
  \label{eq:renormPattern1}
  \mathcal{O}_{p,0,r}^{\mathcal{D}} &=& \sum\limits_{j=0}^{r}\, Z_{r,r-j}\, [\mathcal{O}_{p+j,0,r-j}^{\mathcal{D}}]
  \, , \\
  \label{eq:renormPattern2}
  \mathcal{O}_{p,q,0}^{\mathcal{D}} &=& \sum\limits_{j=0}^{q}\, Z_{q,q-j}\, [\mathcal{O}_{p+j,q-j,0}^{\mathcal{D}}]
  \, 
\end{eqnarray}
and hence have the same anomalous dimensions.

The total derivative basis is useful for connecting continuum quantities to lattice ones in non-perturbative studies, see e.g.~\cite{Gockeler:2004wp, Gracey:2009da}. The mixing matrix is also triangular in this basis ($\gamma_{N,k}^{\mathcal{D}} = 0$ if $k > N$)
and, as was the case for the Gegenbauer basis, the diagonal elements are just the forward anomalous dimensions $\gamma_{N,N}$~\cite{Moch:2017uml}. We can again drop the superscript ${\mathcal{D}}$ due to basis independence. In this basis, the anomalous dimensions for low-$N$ operators are known up to the three-loop level; see {\cite{Gracey:2009da}} for analytical results and {\cite{Kniehl:2020nhw}} for a numerical extension of these. It is also possible to transform the anomalous dimensions in the $\mathcal{D}/\mathcal{G}$ basis to those in the $\mathcal{G}/\mathcal{D}$ one using
\begin{equation}
\label{eq:basisTrans}
    \sum\limits_{j=0}^{N}\, (-1)^j\frac{(j+2)!}{j!}\, \gamma_{N,j}^{\mathcal{G}} \,=\,
    \frac{1}{N!}\, \sum\limits_{j=0}^{N}\,
    (-1)^j\binom{N}{j}\frac{(N+j+2)!}{(j+1)!}\, \sum\limits_{l=0}^{j}\, \gamma_{j,l}^{\mathcal{D}}.
\end{equation}
Note that this is not a 1-to-1 relation between the anomalous dimensions in both bases; the best we can do is relate specific sums to each other.
\section{Constraints on the anomalous dimensions in the total derivative basis}
\label{sec:constraints}
Focusing now on the total derivative basis, it turns out that the elements of the mixing matrices are not all independent. Instead they are subject to particular constraints, 
which define useful relations between them in the chiral limit.
Starting from Eq.~(\ref{eq:partialAct}) we can derive the following relation between the bare operators by acting $N$ times with a partial derivative on $\mathcal{O}_{N,0,0}^{\mathcal{D}}$
\begin{equation}
\label{bareREL}
    \mathcal{O}_{0,N,0}^{\mathcal{D}}-(-1)^N \sum_{j=0}^{N}(-1)^j\binom{N}{j}\mathcal{O}_{j,0,N-j}^{\mathcal{D}} \,=\, 0
    \, .
\end{equation}
Now using the renormalization equations (\ref{eq:renormPattern1}), (\ref{eq:renormPattern2}) and performing some simple algebra, we find a relation between the renormalization factors, and hence between the anomalous dimensions
\begin{align}
\label{mainConj}
    \gamma_{N,k}^{\mathcal{D}} &\,=\, 
    \binom{N}{k}\sum_{j=0}^{N-k}(-1)^j \binom{N-k}{j}\gamma_{j+k,j+k} 
    \nonumber\\&+ \sum_{j=k}^N (-1)^k \binom{j}{k} \sum_{l=j+1}^N (-1)^l \binom{N}{l} \gamma_{l,j}^{\mathcal{D}}
    \, .
\end{align}
As the relation holds a priori at the level of the renormalization constants, the corresponding relation between the anomalous dimensions is valid to all orders in $a_s$. 

Putting now $k=0$ in Eq.~(\ref{mainConj}) yields
\begin{equation}
\label{mainK0}
    \gamma_{N,0}^{\mathcal{D}} = (-)^N\Bigg[\sum_{i=0}^{N}\gamma_{N,i}^{\mathcal{D}}-\sum_{j=1}^{N-1}(-)^j\binom{N}{j}\gamma_{j,0}^{\mathcal{D}}\Bigg].
\end{equation}
This relation allows us to recursively build up the last column of the mixing matrix, provided we can determine the first sum between brackets. We now briefly explain that this is in fact possible. From the renormalization structure of the operators, Eq.~(\ref{eq:renormPattern1}), it is clear that the bare matrix element of $\mathcal{O}_{N+1}$ is related to the sum of renormalization factors $\sum_{i=0}^{N}Z_{N,i}$, and hence its $1/\epsilon$-pole will be related to $\sum_{i=0}^{N}\gamma_{N,i}$. Collecting this information in the quantity $\mathcal{B}(N+1)$ we can then write
\begin{equation}
\label{eq:BtoSum}
\mathcal{B}(N+1) \,=\, \sum_{j=0}^{N} \gamma_{N,j}^{\mathcal{D}}.
    \, 
\end{equation}
Substituting into Eq.~(\ref{mainK0}) leads to
\begin{equation}
\label{g0-from-B}
    \gamma_{N,0}^{\mathcal{D}} \,=\, \sum_{i=0}^{N}(-1)^i\binom{N}{i}\mathcal{B}(i+1).
\end{equation}
This implies that the last column of the mixing matrix can be directly related to a fixed-moment Feynman diagram calculation of matrix elements of operators without total derivatives.

Going back to the general-$k$ relation, Eq.~(\ref{mainConj}), we emphasize that the only assumption made in its derivation was the use of the chiral limit, which imposes constraints on the renormalization structure of the operators. Hence, it can be used as an order-independent consistency check, which any expression for $\gamma_{N,k}^{\mathcal{D}}$ has to obey. 
Alternatively, Eq.~(\ref{mainConj}) allows for the construction of the full mixing matrix starting from the forward anomalous dimensions $\gamma_{N,N}$ 
and the last column $\gamma_{N,0}^{\mathcal{D}}$ at any order of perturbation theory.

So, with partial information being available even to five-loop order, one can produce an ansatz for the off-diagonal elements of the mixing matrix and use Eq.~(\ref{mainConj}) to test its self-consistency. The last column will then serve as a boundary condition.
This leads to a 4-step algorithm for constructing the mixing matrix:

    \textbf{1.} Starting from the bare OMEs in Eq.~(\ref{eq:OMEs}), one determines the all-$N$ expression for the last column entries $\gamma_{N,0}^{\mathcal{D}}$ of the mixing matrix, cf. Eq.~(\ref{g0-from-B}).
    
    \textbf{2.} Next, one calculates a sum of the forward anomalous dimensions
    \begin{equation}
    \label{eq:DiaSum}
        \binom{N}{k}\, \sum_{j=0}^{N-k}\, (-1)^j \binom{N-k}{j}\, \gamma_{j+k,j+k}
        \, .
    \end{equation}
    The structure of the result can then be used to construct an ansatz for the off-diagonal elements.
    
    \textbf{3.} Using the chosen ansatz, one calculates the double sum 
    \begin{equation}
    \label{eq:DoubleSum}
        \sum_{j=k}^N\, (-1)^k \binom{j}{k}\, \sum_{l=j+1}^N\, (-1)^l \binom{N}{l}\, \gamma_{l,j}^{\mathcal{D}}
    \end{equation}
    and collects everything into Eq.~(\ref{mainConj}). This results in a system of equations in the unknown coefficients of the ansatz, subject to the boundary condition that the expression for $\gamma_{N,k}^{\mathcal{D}}$ has to agree with the previously found expression for $\gamma_{N,0}^{\mathcal{D}}$ from step 1.
    
    \textbf{4.} If one finds a unique solution for the system of equations, one has successfully determined the final expression for the off-diagonal elements of the mixing matrix. If such a solution is not found, some terms will remain in Eq.~(\ref{mainConj}). The structure of these remnant terms can be used to adapt the ansatz, leading one back to step 3.
    
During the course of this algorithm, some non-trivial sums appear, cf. Eq.~(\ref{eq:DiaSum}) and Eq.~(\ref{eq:DoubleSum}). These can be evaluated using algorithms of symbolic summation, which are nicely implemented in the {\sc Mathematica} package {\sc Sigma}~\cite{Schneider2004} by Carsten Schneider.
\section{Results}
\label{sec:results}
Using the consistency relation and the algorithm introduced in the previous section, we have calculated the off-diagonal elements of the anomalous dimension matrix up to five-loop order in the leading-$n_f$ approximation. As illustration, we quote the five-loop result
{\footnotesize
\begin{eqnarray}
\gamma^{\mathcal{D}, (4)}_{N,k} &=& \frac{16}{81}n_f^4 C_F\Bigg\{\frac{1}{12}\Big(S_{1}(N)-S_{1}(k)\Big)^4\Big(\frac{1}{N+2}-\frac{1}{N-k}\Big)\nonumber\\&&+\frac{1}{3}\Big(S_{1}(N)-S_{1}(k)\Big)^3\Big(\frac{5}{3}\frac{1}{N-k}+\frac{2}{N+1}-\frac{11}{3}\frac{1}{N+2}+\frac{1}{(N+2)^2}\Big)\nonumber\\&&+\frac{1}{2}\Big(S_{1}(N)-S_{1}(k)\Big)^2\Big(S_{2}(N)-S_{2}(k)\Big)\Big(\frac{1}{N+2}-\frac{1}{N-k}\Big)\nonumber\\&&+\Big(S_{1}(N)-S_{1}(k)\Big)^2\Big(\frac{1}{3}\frac{1}{N-k}-\frac{13}{3}\frac{1}{N+1}+\frac{2}{(N+1)^2}+\frac{4}{N+2}\nonumber\\&&-\frac{11}{3}\frac{1}{(N+2)^2}+\frac{1}{(N+2)^3}\Big)\nonumber\\&&+\Big(S_{1}(N)-S_{1}(k)\Big)\Big(S_{2}(N)-S_{2}(k)\Big)\Big(\frac{5}{3}\frac{1}{N-k}+\frac{2}{N+1}-\frac{11}{3}\frac{1}{N+2}+\frac{1}{(N+2)^2}\Big)\nonumber\\&&+\frac{2}{3}\Big(S_{1}(N)-S_{1}(k)\Big)\Big(S_{3}(N)-S_{3}(k)\Big)\Big(\frac{1}{N+2}-\frac{1}{N-k}\Big)\nonumber\\&&+\Big(S_{1}(N)-S_{1}(k)\Big)\Big(\frac{2}{3}\frac{1}{N-k}+\frac{2}{N+1}-\frac{26}{3}\frac{1}{(N+1)^2}+\frac{4}{(N+1)^3}-\frac{8}{3}\frac{1}{N+2}\nonumber\\&&+\frac{8}{(N+2)^2}-\frac{22}{3}\frac{1}{(N+2)^3}+\frac{2}{(N+2)^4}\Big)+\frac{1}{4}\Big(S_{2}(N)-S_{2}(k)\Big)^2\Big(\frac{1}{N+2}-\frac{1}{N-k}\Big)\nonumber\\&&+\Big(S_{2}(N)-S_{2}(k)\Big)\Big(\frac{1}{3}\frac{1}{N-k}-\frac{13}{3}\frac{1}{N+1}+\frac{2}{(N+1)^2}+\frac{4}{N+2}\nonumber\\&&-\frac{11}{3}\frac{1}{(N+2)^2}+\frac{1}{(N+2)^3}\Big)+\frac{2}{3}\Big(S_{3}(N)-S_{3}(k)\Big)\Big(\frac{5}{3}\frac{1}{N-k}+\frac{2}{N+1}\nonumber\\&&-\frac{11}{3}\frac{1}{N+2}+\frac{1}{(N+2)^2}\Big)+\frac{1}{2}\Big(S_{4}(N)-S_{4}(k)\Big)\Big(\frac{1}{N+2}-\frac{1}{N-k}\Big)+\frac{2}{3}\frac{1}{N-k}\nonumber\\&&-\frac{2}{3}\frac{1}{N+1}+\frac{2}{(N+1)^2}-\frac{26}{3}\frac{1}{(N+1)^3}+\frac{4}{(N+1)^4}-\frac{8}{3}\frac{1}{(N+2)^2}+\frac{8}{(N+2)^3}\nonumber\\&&-\frac{22}{3}\frac{1}{(N+2)^4}+\frac{2}{(N+2)^5} \Bigg\}
\, .
\end{eqnarray}
}
Here $C_F=(n_c^2-1)/(2n_c)$. For more details and for the lower-loop expressions we refer the reader to our main paper \cite{Moch:2021cdq}. 

Furthermore, by transforming our results to the Gegenbauer basis using Eq.~(\ref{eq:basisTrans}), we have an independent check of the results in~\cite{Braun:2017cih}. Finally, our algorithm can also be used to extend the results in the Gegenbauer basis to the four-loop level, in the leading-$n_f$ approximation, the expression for which can be found in~\cite{Moch:2021cdq}.
\section{Conclusion and outlook}
\label{sec:conclusion}
We have studied the renormalization of non-singlet quark operators including total derivative operators, which appear in the OPE analysis of DIS and DVCS. In doing so, we have derived a novel method for calculating the off-diagonal elements of the anomalous dimension matrix, based on the renormalization structure of the operators in the chiral limit. On the one hand, this provides an independent check of previous calculations in different operator bases. On the other hand, we also derive new results, e.g. the five-loop anomalous dimensions in the leading-$n_f$ limit. In our main paper~\cite{Moch:2021cdq}, we also show that the method can be used beyond the leading-$n_f$ limit. Results here include the anomalous dimensions of low-$N$ operators to five loops in full QCD. By performing a scheme transformation to the RI-scheme, these will also become useful in studies of the hadron structure using lattice QCD.

The presented algorithms, i.e. consistency relations in combination with a direct Feynman diagram computation of the relevant OMEs, allow for an automation of the calculations using various computer algebra programs, such as {\sc Forcer} for the 
calculation of massless two-point functions up to four loops and symbolic summation using {\sc Sigma}.

Finally, it should be straightforward to adapt the method to the calculation of mixing matrices for different operators in QCD and to different models all-together. 
These aspects are left for future studies.

\subsection*{Acknowledgements}
This work has been supported by Deutsche Forschungsgemeinschaft (DFG) through the Research Unit FOR 2926, ``Next Generation pQCD for
Hadron Structure: Preparing for the EIC'', project number 40824754 and DFG grant $\text{MO~1801/4-1}$.
%uncomment the following lines to place a figure
%\begin{figure}[htb]
%\centerline{%
%\includegraphics[width=12.5cm]{Fig1}}
%\caption{Plot of ...}
%\label{Fig:F2H}
%\end{figure}

\bigskip

\bibliographystyle{JHEP}
%\bibliography{omebib}

\providecommand{\href}[2]{#2}\begingroup\raggedright\endgroup

\end{document}